# Analysis of information cascading and propagation barriers across distinctive news events

Abdul Sittar[1] · Dunja Mladenić[1] · Marko Grobelnik[1]

**Abstract**
News reporting, on events that occur in our society, can have different styles and structures, as well as different dynamics of news spreading over time. News publishers have the potential to spread their news and reach out to a large number of readers worldwide. In this paper we would like to understand how well they are doing it and which kind of obstacles the news may encounter when spreading. The news to be spread wider cross multiple barriers such as linguistic (the most evident one, as they get published in other natural languages), economic, geographical, political, time zone, and cultural barriers. Observing potential differences between spreading of news on different events published by multiple publishers can bring insights into what may influence the differences in the spreading patterns. There are multiple reasons, possibly many hidden, influencing the speed and geographical spread of news. This paper studies information cascading and propagation barriers, applying the proposed methodology on three distinctive kinds of events: Global Warming, earthquakes, and FIFA World Cup. Our findings suggest that 1) the scope of a specific event significantly effects the news spreading across languages, 2) geographical size of a news publisher's country is directly proportional to the number of publishers and articles reporting on the same information, 3) countries with shorter time-zone differences and similar cultures tend to propagate news between each other, 4) news related to Global Warming comes across economic barriers more smoothly than news related to FIFA World Cup and earthquakes and 5) events which may in some way involve political benefits are mostly published by those publishers which are not politically neutral.

☒ Abdul Sittar
abdul.sittar@ijs.si

Dunja Mladenić
dunja.mladenic@ijs.si

Marko Grobelnik
marko.grobelnik@ijs.si

[1] Jozef Stefan Institute, Ljubljana, Slovenia

## 1 Introduction

News spreading is one of the most effective mechanisms for spreading information. Mainly there are two different ways of obtaining information: 1) observing or engaging in the event in person, for example, listening to a presidential speech, and 2) coming across information that is propagated from multiple channels or publishers which potentially influence the perception of the event they are reporting on and thus can consequently change the recipient's point of view.

Due to globalization, many events from different areas are internationally relevant. Representation of cross-lingual information about an event should be in a unique format and relevant context as this helps people to understand the entire story of current regional and international events that belong to diverse cultures. Information spreading via news is related to information cascading, where publishers decide to write on an event that is already published by another publisher. The result is subsequent news reporting on the same event, starting from the root news article to the last news article on the same event. The concept is commonly used in social media to find a set of subsequent re-shares starting from the root user (Hong et al., 2017).

By analyzing monolingual sets of news articles about an event, we can help in estimating the importance of the event for a specific language which further provides a basis for understanding to what extent an event is important for a country or a region. However, cross-lingual information cascading enables us to find the tendency and interest of each language group for a specific event. The decision on publishing news relies on some certain factors. News has a major impact on decision making for any country, while international news has an impact in terms of international relations and foreign policy as they can change public opinion, which leads towards impacting important decisions for democratic countries (Wu, 1998). In our contemporary society, international news about different events led us to investigate the reasons why news regarding specific events either spread or do not spread to certain geographic areas. Media focuses on specific foreign and regional events based on some certain factors. For instance, spreading of events may tilt toward developed countries such as United States, the United Kingdom, or Russia. Moreover, it may be due to geographical juxtaposition (latitude, longitude) of countries (Wilke et al., 2012). There is a great deal of negotiation between political actors and journalists in news production to enhance their influence on news coverage (Maurer & Beiler, 2018). Therefore political alignment of publishers can more or less impact their coverage of different events. We expect to observe this difference in reporting on sport events, natural disasters or climate changes. An important step which takes place prior to news spreading is news selection of foreign events. Multiple theoretical explanations for this have been presented in the pasts (Wu, 2007; Chang & Lee, 1992).

Two of the determinants for news coverage are economic conditions and association between countries (Chang & Lee, 1992). Cultural values and differences also impact information selection, analysis, and propagation. For instance, if two countries are culturally more similar, there are more chances that there will be heavier news flow between them (Wu, 2007).

This study is focused on three popular events in three different domains: sports, climate change, and natural disasters. The rationals behind selecting these domains is that they are expected to have different influence on the public and differ in information spreading. Natural disasters such as floods, earthquakes, and tsunami waves are unfortunate events caused by the natural and geological processes of Earth. One would expect that news regarding natural disasters is mostly objective. Climate change, such as Global Warming and pollution, is a very controversial topic, with political interests of different actors. Thus the reporting is expected to be selective and biased. Sports news can be considered quite political in nature involving prediction and claims speculating about the results of a game.

This article makes the following contributions:

– Information cascading theory has been adopted to event-centric news analysis.
– We provide new insights into the phenomenon of information propagation in news for different domains.
– With the incorporation of online sources, we are able to retrieve and enhance data related to information barriers.
– We provide the analysis of the influence of multiple barriers on information propagation across distinctive news events.

The remainder of this paper is structured as follows: Section 2 provides related work on information spreading barriers, and an analysis of news events in different domains. In Section 3, we provide details about our data sets and data enhancement. After elaborating on the research methodology in Section 4, the experiments are described in Section 5. Section 6 provides a brief discussion about event-related findings and the corresponding results. Finally, Section 7 presents the conclusion and some ideas for future work.

### 1.1 Hypothesis and research questions

Our research hypothesis states that depending on the nature of an event, there will be variations in information spreading behavior across the observed barriers. We tested our hypothesis on three types of events which are distinct: an earthquake within the natural disaster domain, the FIFA World Cup within the sports domain, and Global Warming within the climate change domain. In order to aid understanding of the influence of different barriers on information spreading, this article set four research questions:

**Q1:** What are the properties (size and ratio) of cascading chains in events of different domains?

**Q2:** Do the different information cascading chains have any relation to each other?

**Q3:** How do economic, geographical, time zone, and cultural values influence news spreading in events of different domains?

**Q4:** What is the correlation of news spreading of events in different domains to the political alignment of the news publishers?

## 2 Related work

The concept of information spreading is a broad topic and has an enormous number of research dimensions. As this study focuses on information spreading and respective barriers, we review six different types of interconnected related works: information spreading and contrasting events, linguistic, economical, geographical and time zone differences as well as political and cultural barriers.

**Information spreading and contrasting events**  As this topic is much known globally and pertinent to cross-cultural studies, various studies have been conducted in the past to understand various aspects of information spreading. Among them are understanding the key features in sports-related information spreading (Alla et al., 2011), temporal aspects while modeling information spreading (Miritello et al., 2011), modeling information diffusion in online social networks (Kumar et al., 2020), and modeling the information dissemination under disaster (Cui et al., 2020). Moreover, our study focuses on modeling events in different domains (sports, natural disasters, and climate change) and understanding hidden patterns in the flow of information. Our preliminary results show that events have different spreading patterns depending on the domain; in particular, we notice the spreading of news related to natural disasters comes across fewer barriers than the news related to sports as well as news related to climate change.

**Linguistic Barriers** Much of the research on linguistic barriers has focused on understanding properties of information propagation such as speed, size, and structure. Some researchers focus on cross-lingual information diffusion to understand information cascading (Jin, 2017) in social networks. There are attractive options on social media such as share or retweet that have been used mostly to understand information cascading rather than to understand the semantic meaning of the text. Unlike within the social media domain, finding cross-lingual similarity in the context of news is one of the top priorities. There are numerous studies conducted on semantic textual similarity in the literature. The concept of measuring semantic textual similarity is based on estimating the semantic relatedness between two or multiple texts (Glavaš et al., 2018). Multiple techniques have been developed to estimate cross-lingual semantic similarity of texts achieving satisfactory results, such as word-to-word translation (Vulic & Moens, 2014), dictionary based translations (Krajewski et al., 2016) and word embeddings methods for semantic similarity (Şenel et al., 2017). These approaches are mostly evaluated in the context of plagiarism detection and discourse analysis. In our case, we are focusing on understanding whether a piece of news is discussing something related to an event or not. Therefore, we consider concept-based similarity more pertinent and calculate similarity based on Wikipedia concepts[1]. The existing research on information cascading focuses on social networks and relies on social media features (e.g., retweets, share). Our method presents a new approach to cascading based on news spreading. Firstly, we observe information flow in mono- and cross-lingual settings. Secondly, we go beyond information flow based on textual similarity and show the flow of news related to events in different domains and in different languages from the point of view of temporal elements (e.g., monthly spread of news). Moreover, language as a part of cultural values and beliefs, largely emerges to have a worldwide significant role in news selection regarding different domains and their propagation at different times. Therefore, cross-lingual news can help in understanding differences between high-resource languages and low-resource languages in the process of news creation and propagation.

**Economical Barriers** According to news flow theories, multiple determinants impact international news spreading. The economic power of a country is one of the factors that influence news spreading. Moreover, economic variations have different influence for different events (e.g., protests, conflicts, disasters) (Segev, 2015). The magnitude of economic interactivity between countries can also impact the news flow (Wu, 2007). Economic growth/income level shows the economic condition of a country. Multiple organizations are working on generating prosperity and welfare indexes on a yearly basis. Among them, "The

---

[1] http://wikifier.org/info.html

Legatum Prosperity Index" and "Human Development Index" are popular[2],[3]. These prosperity indexes are already used to compare and draw prosperity relations within a country or between the countries to understand different aspects such as education, business infrastructure, and technology (Büyüksarıkulak & Kahramanoğlu, 2019). Our intention here is to compare news spreading in relation to the income level in the country of the news publisher.

**Geographical and Time Zone Barriers** Geographical representation of entities and events has been utilized extensively in the past to detect local, global, and critical events (Quezada et al., 2015; Wei et al., 2020; Watanabe et al., 2011; Andrews et al., 2016). It can help us to observe the proximity effects on corresponding research questions. It has been said that countries with close distance share culture and language up to a certain extent which can further reveal interesting facts about shared tendencies in information spreading (Segev & Hills, 2014; Segev, 2015). Our motivation for using geographical locations is to analyze the impact of geographical proximity on news spreading in different domains. In addition, to represent relative time, the time zone is an alternative to geographical difference between countries (Dagon et al., 2006). The publishing time of news articles has a strong association with time zones; therefore, our analysis will also take it into account.

**Political Barriers** News agencies tend to follow the national context in which journalists operate. One of the related examples is the SARS epidemic study which found that cross-national contextual values such as political and economic situations impact the news selection (Camaj, 2010). It will be true to say that fake news is produced based on many factors and it is surrounded by a paramount factor that is political effect. A great amount of work regarding fake news dwells on different strategies, while few studies considered political alignment to have a compelling effect on news spreading (Bakshy et al., 2015; Maurer & Beiler, 2018). Maurer and Beiler (2018) strongly proved it to be a major strategy in news agencies to control the news and change accordingly due to the involvement of journalists and political actors. One can expect that news on climate change and sports is more prone to be altered due to the political alignment of publishers than news on natural disasters. Our study incorporates political alignment of news publishers.

**Cultural Barriers** Countries that share a common culture are expected to have heavier news flow between them when reporting on similar events (Wu, 2007). There are many quantitative studies that found demographic, psychological, socio-cultural, source, system, and content-related aspects (Al-Samarraie et al., 2017). There are many models that have tried to explain cultural differences between societies. Hofstede's national culture dimensions (HNCD) have been widely used and cited in different disciplines (Khosrowjerdi et al., 2020; He & Lee, 2020). It has also been criticized by many researchers for reducing culture into dimensions. Originally, HNCD were comprised of four dimensions: power distance, individualism, uncertainty avoidance by individuals, and masculinity vs. femininity. It then further extended by two dimensions: long-term vs. short-term orientation and indulgent versus restrained. Scores for all dimensions for different countries have been presented in Table 8. Following is a description of these dimensions.

---

[2]http://hdr.undp.org/en/content/human-development-index-hdi
[3]https://www.prosperity.com/

**Power distance index:** shows the extent to which the power inequalities in society agree with each other, such as children and parents, youths and elders, students and teachers. A higher degree of the index indicates that power is distributed transparently, whereas a lower degree of the index is a sign that people question authority.

**Uncertainty avoidance by individuals:** relates to the degree to which a society is tolerant toward unknown, unusual, and novel situations. Higher score signifies that people prefer to choose stiff codes of behaviors, and rules etc., whereas a lower score indicates that society prefers to impose fewer regulations.

**Non-individualistic cultures:** Individualism is characterized by loose ties between individuals and a focus on privacy and personal integrity. Higher score means that society, people are more integrated into groups. On the contrary, a lower score shows an emphasis on individualism.

**Masculinity vs. Femininity:** This describes whether a society is dominated by masculine culture; in this case, the society is assertive and competitive with a high level of gender inequality. Alternatively, in the case where the society is oriented more towards femininity, there prevails a modest and caring values. Higher score means society is more dominated by masculinity and vice versa.

**Long-term orientation:** These societies are likely to emphasize savings and hardworking behaviors preparing for potential critical events in the future. On the contrary, short-term oriented societies do not focus on a futuristic approach. A higher score indicates that a society is more future oriented whereas lower score signifies that a society places a greater emphasis on ancestors and honors traditions.

**Indulgence vs. Restraint:** HNCD claims that indulgent societies focus more on personal fulfillment and jolly behavior but restrained societies focus on having personal wishes and happiness controlled by social norms. In this dimension, higher index indicates a higher degree of freedom in fulfilling the human desires in a society. On the other hand, lower index indicates that a society controls the gratification of needs.

There are fewer studies that take cultural values into account while recognizing the flow of news related to different events. Our study considers cultural values to be a significant factor in the spread of cross-lingual news across different countries. For illustration of this effect, we show the score of each dimension for 67 countries in Table 8.

## 3 Data description and pre-processing

This section presents the dataset we have collected for the purpose of our research, definitions of the basic terms (Information Propagated, Unsure, and Information not propagated), and enhancements of the dataset.

### 3.1 Data description

We have created a corpus that consist of 7773 news articles published between 2015 – 2020 in one of the five selected languages (English, Portuguese, German, Spanish and Slovenian) (Sittar et al., 2020). The cross-lingual news articles were represented as vectors of Wikipedia concepts obtained by annotating articles using the Wikifier Service[4]. As in bag-of-words representation, we calculate tf-idf weight for each element of the vector, which in our case were Wikipedia concepts. To identify information propagation, we measure similarity between each article and all the articles published later than the article using the feed-forward mechanism. To measure similarity between two articles, we calculate cosine similarity of their vectors (similarity scores varies between 0 and 1. 0 means minimum and 1 means maximum similarity). The number of final pairs was 7817. We have published this in a paper entitled "A Dataset for Information Spreading over the News"[5] created with the aim to analyze propagation of information over news articles. All the articles in our dataset belong to one of the three kinds of events: (1) Global Warming, (2) FIFA World Cup, and (3) Earthquake (see Table 1). Cross-lingual similarity between the news articles was calculated based on Wikipedia concepts, as each article was represented by the associated Wikipedia concepts (Brank et al., 2017) as provided by the Event Registry systems (Leban et al., 2014), which we have utilised for collecting the articles. Table 1 shows statistics for each dataset. Based on the similarity score, pairs of articles are classified into one of the following classes:

**Information Propagated:** These articles are likely to discuss a similar event.

**Unsure:** There is uncertainty whether the information is propagating or not.

**Information not Propagated:** These articles are likely to involve a discussion about different events.

### 3.2 Dataset enhancement

To understand the effect of multiple barriers, we have enriched the dataset using information related to each barrier under observation, obtaining the data from different sources as described in the rest of this section. For most of the barriers (Economic, Geographical, Time Zone, Political, and Cultural), we have utilised information connected with the country of the publisher of the news articles.

### 3.2.1 Economical data

Deciphering information about economical barriers across different countries depends upon the economic profile of countries. We collected economic profiles of all the countries using the latest prosperity ranking 2019[6] and income levels (High, Upper Middle, Lower Middle, and Low-income level) using World Bank Country data[7]. Data along with the economic profiles of the countries of the news publishers can be found on the Zenodo repository (version 1.0.2)[8].

---

[4]http://wikifier.org/info.html
[5]https://zenodo.org/record/3950065
[6]https://www.prosperity.com/rankings
[7]https://datahelpdesk.worldbank.org/knowledgebase/articles/906519-world-bank-country-and-lending-groups
[8]https://zenodo.org/record/4117411

**Table 1** Statistics about dataset showing for each event type, the number of articles per language

| Dataset | Event type | Articles per language | | | | | Total articles |
|---|---|---|---|---|---|---|---|
| 1 | FIFA World Cup | Eng | Spa | Ger | Slv | Por | 2682 |
| | | 983 | 762 | 711 | 10 | 216 | |
| 2 | Earthquake | 941 | 999 | 937 | 19 | 251 | 3147 |
| 3 | Global Warming | 996 | 298 | 545 | 8 | 97 | 1944 |

#### 3.2.2 Linguistic data

For the linguistic barrier, the dataset already encompasses information on the natural language in which the article is written. As the articles are represented in a language neutral way - by Wikipedia concepts, we are able to compare articles in different languages and in the case of high similarity, we assume that the information is spreading from older to newer articles. Data regarding the linguistic analysis for all three events can be found on the Zenodo repository (version 1.0.2).

#### 3.2.3 Political data

To understand the influence of political actors, we have obtained profiles of publishers from the Wikipedia info-box[9]. However, the profile of some of the publishers did not exist on Wikipedia (see Fig. 1), as a result of which the number of articles was reduced after excluding those with missing publishers' profiles. Table 2 shows the total number of publishers with profiles and a reduced list of articles. The purpose of this publishers' profile was to understand the political alignment of publishers for event-specific news articles. Data regarding political barrier analysis for all three events can be found on the Zenodo repository (version 1.0.2).

#### 3.2.4 Geographical data

For geographical analysis of events, there was a need for the physical location of the publishers' headquarters or the geographical location of the country to which the publisher belongs. Publishers' profiles (see Section 3.2.3) help us to determine the name of the headquarters, which is then used to obtain the location (latitude/longitude) of a publisher's country. We grouped the newspapers according to their locations to see the distribution of news publishers over world map. We also grouped the news articles for each country to see their distribution over world map. Data regarding geographical barrier for all three events can be found in Zenodo repository.

#### 3.2.5 Time zone data

Having obtained the location of a publisher from its headquarters (see Section 3.2.4) and knowing that the time zone is associated with the geographical location, we fetched the general time zone of all the countries for all the publishers. In case of a country having multiple time zones, we selected one of them randomly. Data regarding time zone barrier for all three events can be found in the Zenodo repository.

---

[9]https://en.wikipedia.org/wiki/Help:Infobox

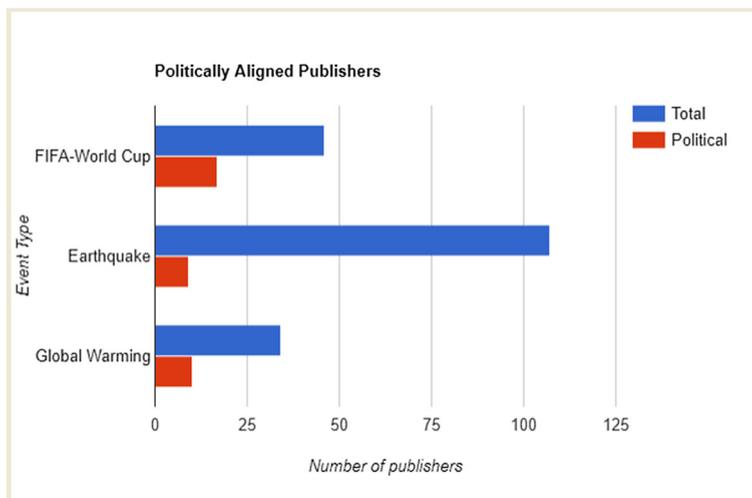

**Fig. 1** Number of publishers with and without political alignment

#### 3.2.6 Cultural data

To study the cultural barrier, we collected six dimensions showing cultural values of each country as suggested in Khosrowjerdi et al. (2020). Cultural dimensions have been assembled by an IBM study on different international populations and by different researchers (Hoftede et al., 2010). We extracted the list of countries along with these dimensions (shown in Table 8) from http://geerthofstede.com/research-and-vsm/dimension-datamatrix/. Table 8 shows the values of cultural dimensions. As only a few countries were missing from this list, we have excluded them from our data sets. Data regarding cultural barrier for all three events can be found in Zenodo repository.

## 4 Methodology

The presented research focuses on analysis of information propagation in news across different barriers in different domains. To this end, we propose a novel methodology consisting of several steps, as shown in Fig. 2.

**Table 2** Statistics on available Wikipedia profiles for publishers and the number of corresponding news articles in our dataset

| Domain | Publishers' profiles | Total articles |
| --- | --- | --- |
| FIFA World Cup | 515 | 324 |
| Earthquake | 341 | 406 |
| Global Warming | 399 | 226 |

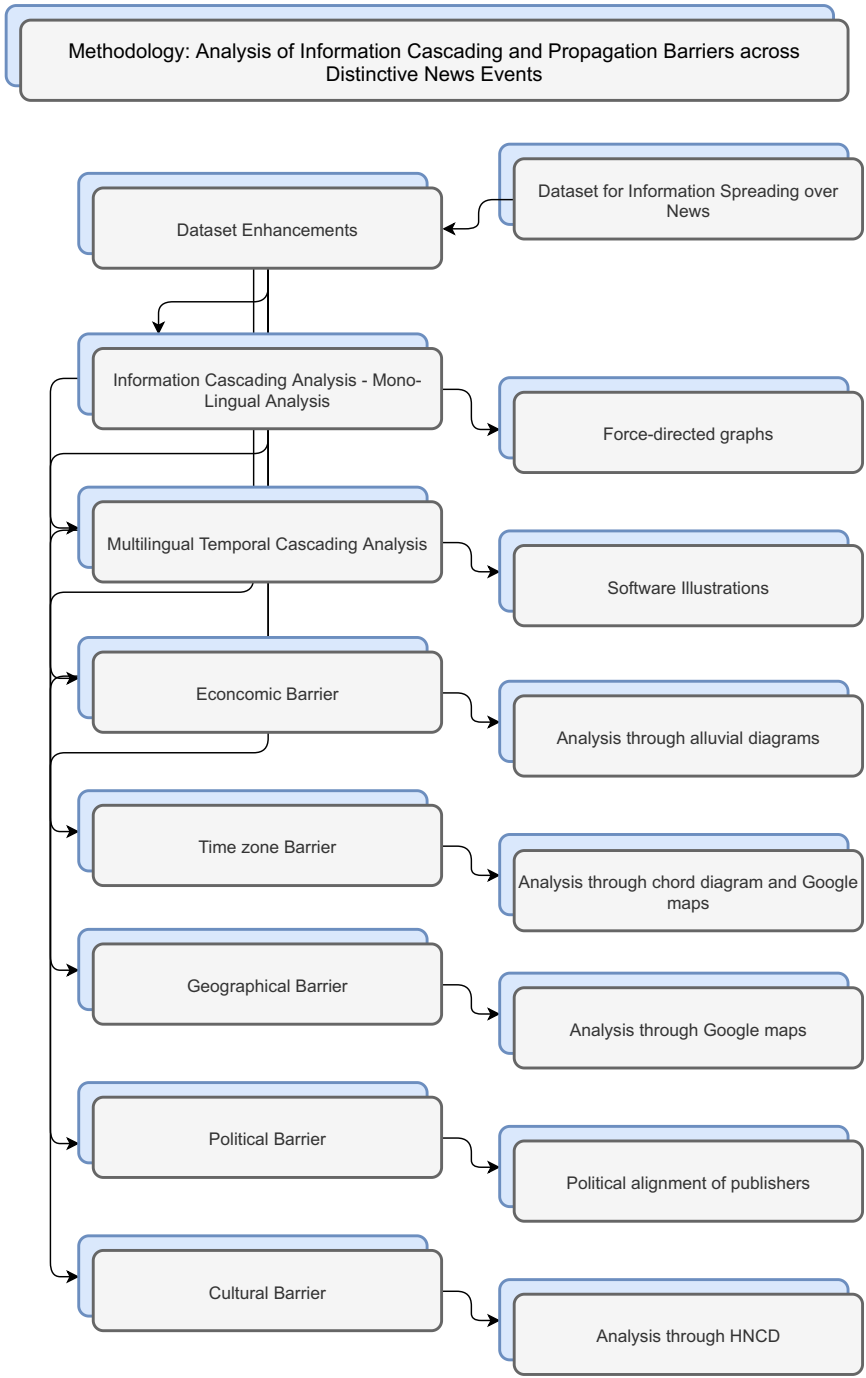

**Fig. 2** Methodology to analyze information cascading and information propagation barriers. The dataset containing pairs of news articles exhibiting information spreading is the initial input. The data is enhanced to include background information on different barriers; analysis is performed to gain insights into information cascading and information spreading across different barriers

In the first step, we perform data enhancement to incorporate information related to different barriers as described in Section 3.2. Then we provide a distinct visualization for each of the observed barriers to help in the understanding of the direction and intensity of information spreading. The rationals behind each of the visualization methods are provided in the next part of this section. We perform different type of analysis mentioned in the proposed methodology on three types of events that are distinct in nature. We select sports, climate change, and natural disasters as three different domains with the rationale that each of these domains has a different influence on the general public. We utilise an existing dataset that is the result of our previous work (Sittar et al., 2020). For the purpose of the presented research, we perform some necessary enhancements of the original dataset in order to expand the focus beyond linguistic barriers.

The main aim of our study is to analyze multiple influences on the news spreading in three types of events (earthquakes, FIFA World Cup, and Global Warming) belonging to three different domains. We focus on information cascading and cross-lingual information spreading across geographical, economical, time zone, political and cultural barriers. For linguistic cascading and cross-lingual information spreading, the existing dataset already encompasses the needed information. For all other barriers, we collect additional information as described in Section 3.

For analysis of information spreading in mono-lingual settings, we perform network analysis using force-directed graphs (see Fig. 4). Force-directed graphs help to visualize connections between objects in a network and use to uncover relationship between the objects. Pairs of news articles with cosine similarity are passed as input to generate multiple chains of news articles as output (Similarity scores varies between 0 and 1, where 0 indicates no similarity and 1 indicates maximum similarity). For instance, if there exists two pairs with the news articles (A, B) and (B, C) respectively, then they will produce a chain/community. In the proposed visualization, we use a different color for each language. First we construct a network for each language for each event and then identify the chains/communities within a language. To detect the communities in each network, we use the Girvan-Newman algorithm based on Edge-Betweenness modularity. This algorithm takes a graph/network as input and provides possible communities along with different nodes. To analyze the communities, we looked into the text of news articles with communities of different size. The detailed implementation of the mono-lingual analysis using force directed graphs is available on Github[10].

For multi-lingual temporal information cascading, a visualization has been developed using Processing IDE to portray the temporal spread of a list of news articles about each event (see Fig. 6). Processing IDE[11] is a flexible software sketchbook that is used for prototyping new interfaces and services. It has been used in research laboratories of famous companies like Google and Intel for prototyping interfaces and services. It has also been used to visualize the data. For example, Yahoo! and Nokia used it for visualization and New York Times Company R&D Lab used it to visualize the way their news stories travel through social media. We used this software to develop a prototype to visualize the way the news propagates[12]. Similar to network analysis, we pass the pair of news articles as input and time as an additional feature where time provides the month of publishing. This approach enables us to identify the most influential languages about an event and results in an overview of the cascading structure. For instance, we have pairs of cross-lingual news articles along with the cosine similarity (see 3.1). This visualisation tool uses these pairs as input, draws spirals/circles of articles published within the same month and links the news articles that propagate information from one month to another. This enables finding how much information propagates across languages and over time. Figure 3 shows the temporal spreading of information in multi-lingual settings. The time interval captured by our datasets is 2015 – 2020 inclusive (6 years = 72 months). Each dot represents a news article. Each spiral/circle of dots indicates articles that were published within the same month. The connection between the two dots indicates that one article is spreading news to other article. We show two types of connections: within a month and within the following month. If two dots are connected within the same spiral/circle, this means that these two articles are propagating news from one article to other and were published within the same month. If a connection is made from one circle to another circle, it means that the two articles are spreading news and published in two consecutive months (see Fig. 3). The color of each dot represents the language of an article (Find the video file of prototype on Github[13]. The main purpose of the proposed multi-lingual temporal spreading visualisation is to show the overall prospect of information spreading on different events in different languages with time.

---

[10]https://github.com/abdulsittar/Mono-lingual-Analysis
[11]https://processing.org/overview/
[12]https://nytlabs.com/projects/cascade.html

To understand the effect of economical barriers on information propagation, we perform the analysis based on economical categories of different countries of news publishers using the alluvial diagrams for each event (see Fig. 7). Alluvial diagrams are basically flow diagrams that help to discover change in large complex networks. We use the category of each news article (High-Income, Lower-Middle-Income, Upper-Middle-Income, or Low-Income) as input and generate an alluvial diagram for each event. This enables us to count the propagation among different types of economies and find associations between categories for each event.

To capture the similarities and differences across different countries regarding information propagation that is caused by different time zones, we construct a chord diagram for each event (see Fig. 9). A chord diagram is a graphical method to display the interrelationship between data points in a matrix. We pass the UTC time zone and UTC time difference for a pair of news articles as input and generate a chord diagram as output. This provides information regarding important time zones for each event.

To visualize information spreading paths from one country to another, we utilize Google maps (see Figs. 8, and 10). It enables to visualize entities on geographical landscapes. As input to the visualization service we provide the list of country names along with the number of articles originating from the country. On the generated maps, we draw links between the countries based on the intensity of information propagation. It enables to understand the geographical impact on news spreading for events. The map visualization further enables to obtain insights about economic, time zone and cultural relations among countries based on information propagation.

In the political barrier, we categorize the news publishers based on their political alignment (see Table 6). This enables to find the inclination of the news publishers altogether toward the events in different domains based on their political alignment. This also helps to know the political alignment of all the publishers that spread news for instance related to sports or natural disasters.

---

[13]https://github.com/abdulsittar/ProcesssingSketch/blob/master/FIFAWorldCup.mp4

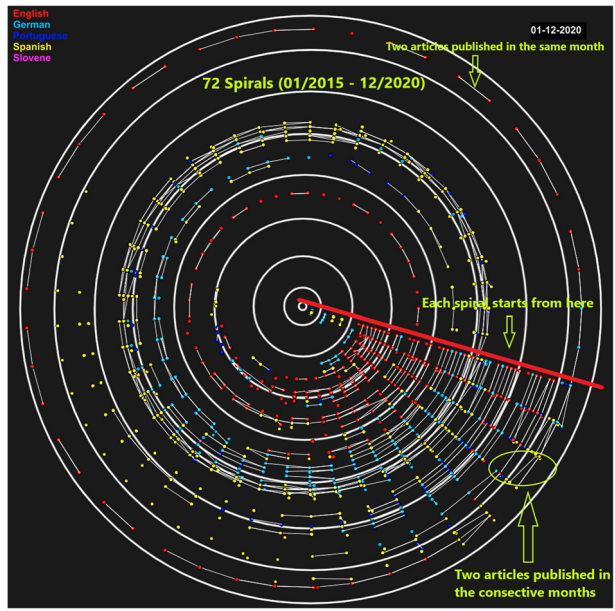

**Fig. 3** Visual depiction of multi-lingual temporal propagation for FIFA World Cup enlarged from Fig. 6

Our proposed methodology uses two ways to approach the analysis: using visualization tools providing diagrams and maps to understand the influence of different barriers. This helps us in answering the last two research questions (see Section 1.1); the second way for approaching the analysis, which is new, provides a mechanism with which we are able to analyze the information cascading into mono-lingual settings and multilingual temporal cascading. This mechanism takes time into account and enables us to visualize the temporal spreading of information through news articles. This furthermore assists us in obtaining insight into the first two research questions).

## 5 Experiments

### 5.1 Propagation analysis through cascading

#### 5.1.1 Mono-lingual analysis

For each event (i.e. Global Warming, earthquakes, and FIFA World Cup), we generated and analyzed networks of similar news articles. The network graph representing these three events is shown below (see Fig. 4). Nodes represent the articles and the color of the node uniquely identifies the language. Although the connected articles are clearly visible, the diagram is too dense to understand fully. Thus, as the next step in the analysis, we fetched important communities (small sub-networks with the largest path from one article to another) in order to analyze the flow of information. A community consists of nodes and edges and basically segregates a group of similar nodes from dissimilar groups (Raghavan et al., 2007). Figure 4 illustrates communities of the news articles that are spreading

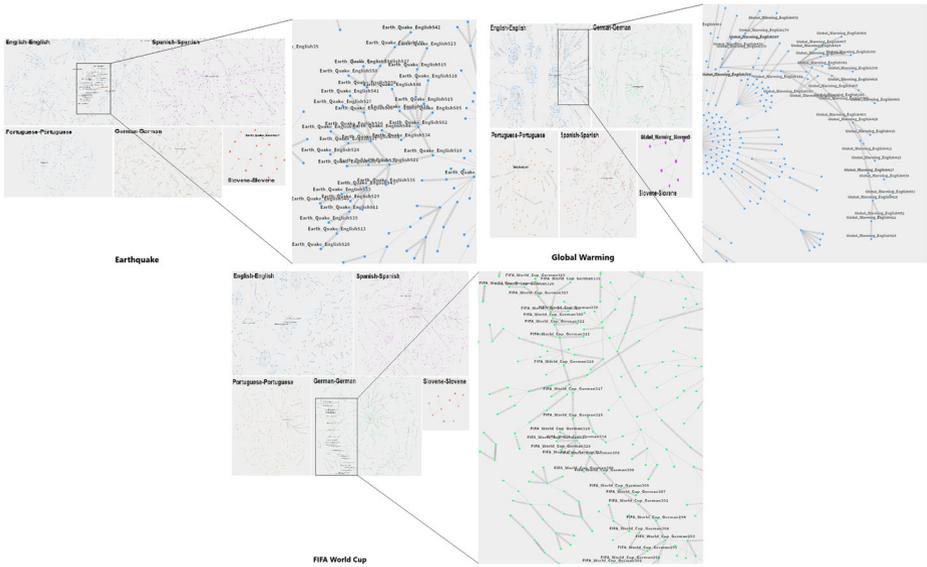

**Fig. 4** Overview of longest cascading chains in different languages related to events such as cascading chains in English, English and German for Earthquake, Global Warming and FIFA World Cup

information from one article to another for all three types of events. We used the Girvan-Newman algorithm based on Edge-Betweenness modularity for detecting communities in networks (Estrada, 2011). To analyze the communities we looked into the text of news articles within the small and large communities. Normally communities with a small number of articles (two or three) contain articles that are copy of each other or are reporting about the same event and the time difference between the articles is small (usually less than 1 hour). For instance, we randomly selected a community/chain of three news articles in English language related to FIFA World Cup domain.

Each of these news articles reports about anonymous attacks on 2022 Qatar World Cup and published within 10 minutes by three different publishers. A community with more articles could involve different discussions than a few articles, therefore we focused on large communities/chains.

### 5.1.2 Mono-lingual propagation - network analysis

For the FIFA World cup, total communities were 147. Each language English, German, Spanish, Portuguese, and Slovene had 47, 67, 26, 6, 0 chains respectively. For each of these languages, the largest communities consisted of 8, 24, 8, 5, 0 news articles whereas small communities consisted of 3 news articles (See Fig. 5). Time difference in the largest community (a chain of 24 news articles in German language) was of almost 3 months (31/08/2016 - 29/11/2016). In this community we see that it is related to the 2018 FIFA World Cup. Generally, by reading the articles manually, we found that most news articles report on the top three matches (Portugal vs. Spain, Egypt vs. Uruguay, and Morocco vs. Iran) minute-wise. In a group of 24 news articles, more than half of the articles were an exact copy of each other and the remaining articles were different only with regard to the varying amount of commentary text. For example, two articles were an exact copy of each other but the second one contained commentary up to the last 9 minutes of the game. Similarly, the other two articles were spreading the same information but the second one also contained extra text praising Cristiano Ronaldo.

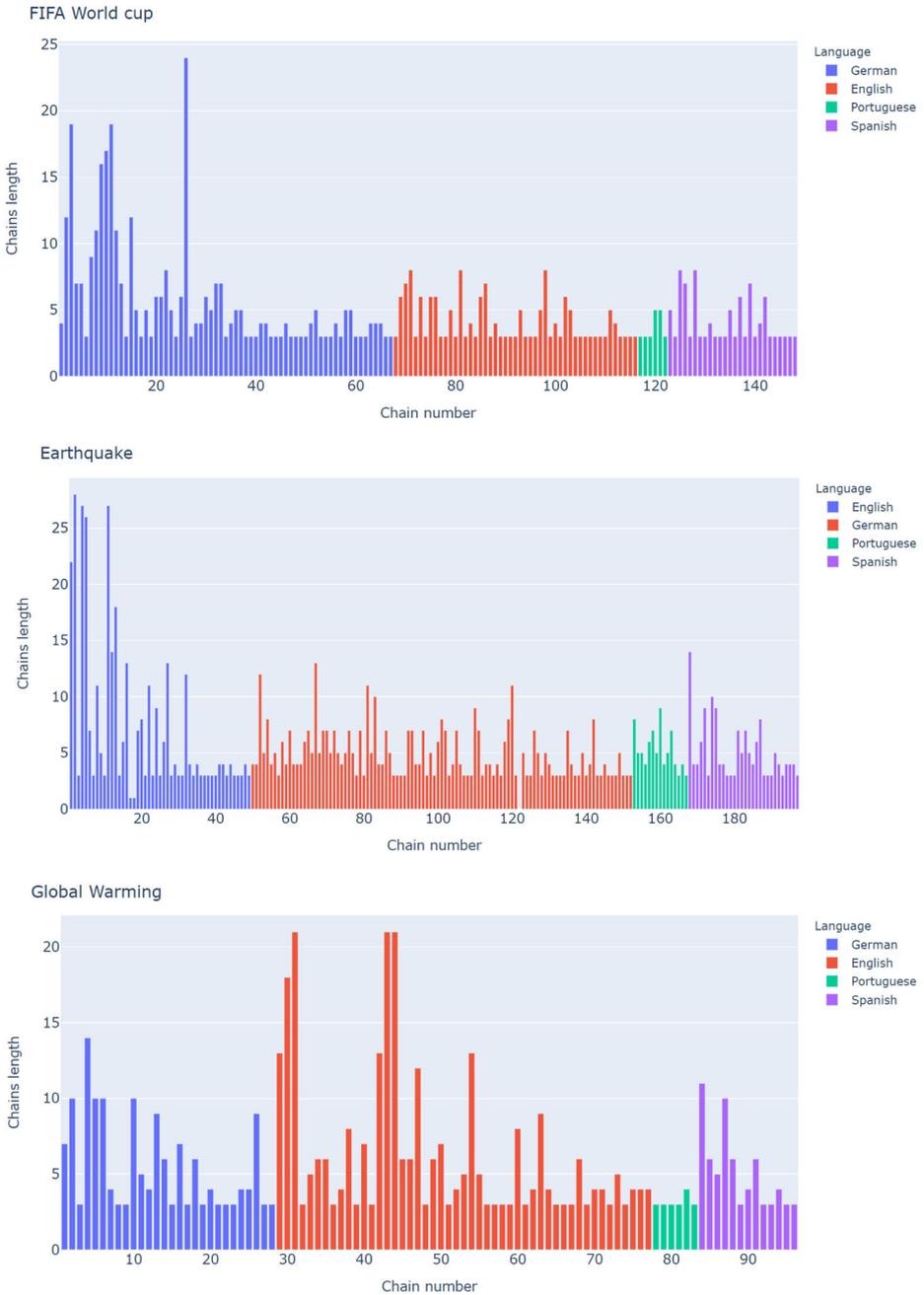

**Fig. 5** Overview of length of chains for Earthquake, Global Warming and FIFA World Cup in different languages

On the other hand, with regard to earthquakes, there were 196 total chains. Each language English, German, Spanish, Portuguese, and Slovene had 49, 101, 31, 15, 0 chains respectively. For each of these languages, the largest chain consisted of 28, 12, 10, 9, 0 news articles whereas small communities consisted of 3 news articles (See Fig. 5). Time difference in the largest chain (a chain of 28 news articles in English language) was of almost 2.5 years (30/12/2017 - 29/04/2020). We see that the discussion within news articles was quite diversified. In general, we have seen that many of the news articles were related to COVID-19 but showed some pertinence to earthquakes. For instance, an article discussed that unlike cyclones or earthquakes, there are no natural disaster constraints in the COVID-19 pandemic situation. Similarly other articles enlightened related aspects such as NCCA (National Commission for Culture and Arts - Philippines) efforts to fight the spread of COVID-19, victims of earthquake, description about community-based disaster management plans in floods, earthquakes, and COVID-19 by German panchayats; the conversation about taking steps for COVID-19 as Tokyo has shockproof buildings for earthquakes; analysis of the budget of the Japanese government for massive earthquakes and tsunamis in earlier years; appraisal of Cuba's medical aid to other countries in time of earthquake and COVID-19 pandemics; and questions comparing which disaster (earthquake, bomb, tornado) could cause death toll. Secondly, out of a group of 28 articles, 8 news articles were found to have a discussion which was about Nasdaq insurance models and risk modeling services. Nasdaq is currently focusing on natural catastrophes with a model spanning to earthquakes, hurricanes, floods, and a number of other perils. Thirdly, four articles were related to earthquakes that occurred in California, New Zealand and Pakistan. Finally, a small number of articles involved unrelated discourses but were connected with earthquakes. For example, the first article was about the strange behavior of cats before an earthquake, second explains an earthquake felt by an English football team, and third article provides the detail about the drop in sales of Japanese auto manufacturer after a massive earthquake.

For Global Warming, there were 95 total chains. Each language English, German, Spanish, Portuguese, and Slovene had 48, 28, 13, 6, 0 chains respectively. For each of these languages, the largest chain consisted of 21, 14, 11, 4, 0 news articles whereas small communities consisted of 3 news articles (See Fig. 5). Time difference in the largest chain (a chain of 21 news articles in English language) was of almost 1 day (30/03/2018 - 31/03/2018). Contrary to the earthquakes, the largest community of news articles regarding climate changes had numerous discussions. Generally, every article was related to Global Warming but involved a variety of discussions ranging from the effect of climate change on COVID-19 to the design of buildings which are suitable for various climate changes. Very few articles were fully similar, with every item of news portraying a different point of view, described from a different perspective. Only three news articles were exactly similar that explain Green recovery is necessary to revive the global economy, whereas three articles comprised of three topics: first climate strike, limiting Global Warming by Donald Trump, and a speech about African actions to combat Global Warming in the United States. Since COVID-19 was affected by climate change, few articles appeared to have WHO related declarations and advice such as UN Chief Antonio Guerres address G20 regarding the fact that major developing and emerging economies together account for 80 percent global emissions, the fact that online streaming sites would not be allowed to upload a video that does not agree with UN intergovernmental panels position regarding exaggerated claims about global warming, and the fact that WHO-NASA affirmed that 2019 was the warmest year ever. The longest communities show the scope of an event for a specific language. Since English is currently the most widely used international language and events such as Global Warming and earthquakes incorporate news internationally, English appeared to have long cascading chains for both events. FIFA World Cup, being held mostly in Europe, has long cascading chains in the German language.

### 5.1.3 Multi-lingual temporal spreading

We observe information spreading via news articles that are published over some time period possibly crossing language barriers. We have built a visualization prototype to show this spreading over the period of six years focusing on information spreading from one language to another language (see Fig. 3)[14].

We have manually selected the articles which are spreading information to some other articles. Among them, we have chosen one chain for each event randomly to check the discourse of articles making the chain (see Fig. 3). Tables 3, 4 and 5 show the titles of the chains of articles with temporal information in each event. To understand the whole discourse, we manually read these chains.

Within the FIFA World Cup domain (see Fig. 6), in the chain of five articles (see Table 3), the first article reports on the FIFA World Cup player Norman Hunter, who was in the hospital due to COVID-19. This article also praised his past victories. The next article published on the same date that shows only upcoming matches schedules, and quiz about statistics of the matches and the names of famous players. This article was related to the FIFA World Cup but is entirely different from the first article. However, contrary to this, the remaining three articles were an exact copy of the first article. According to the dataset, the publishing time varied 3, 5, and 24 hours for the next three articles.

In case of climate changes (see Fig. 6), the chain of five news articles (see Table 4) which were published within 3 days appeared to have both similar and different types of discussions, for example the first and second article reports a general description of the Global Warming phenomenon and how the situation has worsened. Mainly both articles referred to the current analysis that is clear and removes any contradictions about the future threat of Global Warming. There was almost 13 hours of difference in publishing time. The next two articles were seen to have a discussion about Canadian jackets but a single word Global Warming appeared in that context. The publishing time difference were of approximately 3 days. The last article in this chain was published after 5 hours after the second last article. It explains the criticism of the measures that were taken by authorities in order to make climate policy and mentions a protest of students that was recorded and signed by professors to emphasize the need of change for future generations.

Finally, the chain of articles (see Fig. 6) present information spreading regarding an earthquake (see Table 5). It involves two types of discussions where the first article is written about a music festival that explains everything unrelated to earthquakes whereas the other three articles are reporting on an earthquake in Athens. The three articles had publishing time differences of about 10 minutes.

Looking at visualizations of multi-lingual temporal spreading of information, the significant findings are as follows:

1. Almost each year, FIFA World Cup had intensive information propagated in English and Spanish.
2. In case of Global Warming, looking at news articles in the period 2017 – 2020, information was propagated mostly in the English language.
3. For earthquakes, Spanish had the most number of articles compared to the other languages.

---

[14]https://github.com/abdulsittar/ProcesssingSketch

**Table 3** An example of discourse along with publishing time taken from visual propagation (Fig. 6) related to FIFA World Cup

| Publishing Time | Title |
| --- | --- |
| 2019/11/30 - 18:57:00 | Botond Barath and Johnny Russell learn potential group stage opponents at UEFA Euro 2020 |
| 2019/11/30 - 19:24:00 | Euro 2020 Draw: Seeding and Schedule of Dates for Group Fixtures |
| 2019/11/30 - 22:41:00 | Roberto Mancini Says Italy 'Are Not the Favourites' After Euro 2020 Draw |
| 2019/11/30 - 23:29:00 | St. Petersburg fully ready to host Euro-2020 games – authorities - Russia News Now |
| 2019/12/31 - 20:21:00 | Spurned by Neighbors- Qatar Aims for Self-Sufficiency |

**Table 4** An example of discourse along with publishing time taken from visual propagation (Fig. 6) related to Global Warming

| Publishing Time | Title |
| --- | --- |
| 2019/01/10 - 19:31:00 | Die Ozeane heizen sich schneller auf als gedacht |
| 2019/01/11 - 07:46:00 | Die Ozeane heizen sich schneller auf als gedacht - derStandard.at |
| 2019/01/11 - 18:21:00 | Gans dicke, Land Rover zum Anziehen |
| 2019/01/14 - 08:15:00 | Land Rover zum Anziehen,Land Rover zum Anziehen,Süddeutsche Zeitung |
| 2019/01/14 - 13:57:00 | Offener Brief: HNEE-Studierende fordern schnellstmöglichen Kohleausstieg |

**Table 5** An example of discourse along with publishing time taken from visual propagation (Fig. 6) related to Earthquake

| Publishing time | Title |
| --- | --- |
| 2019/06/04 - 17:54:00 | Dub Inc e Horace Andy atuam em julho no festival Musa na praia de Carcavelos |
| 2019/07/19 - 13:13:00 | Terremoto magnitude 5.3 abala Atenas - ISTOÉ Independente |
| 2019/07/19 - 13:26:00 | Terremoto magnitude 5.3 abala Atenas - ISTOÉ DINHEIRO |
| 2019/07/19 - 13:34:00 | Terremoto de magnitude 5.3 abala Atenas,Terremoto de magnitude 5.3 abala Atenas - Alô Limeira! |

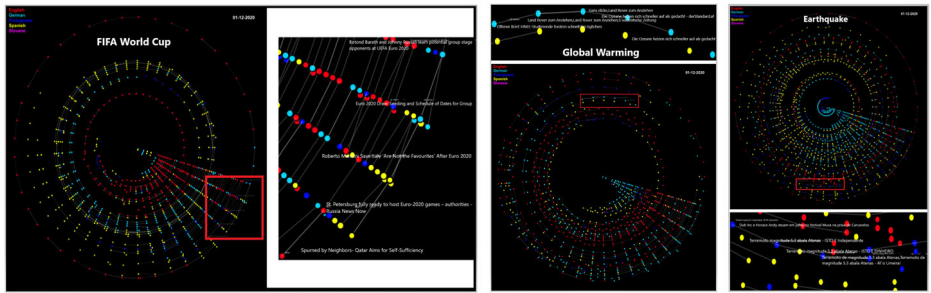

**Fig. 6** Visual depiction of multi-lingual temporal propagation related to FIFA World Cup, Global Warming, and earthquake events

## 5.2 Propagation analysis across other barriers

### 5.2.1 Economical barriers

Propagation of the number of articles from one country to another country and the economic condition of countries has been analyzed. Figure 7 indicates the propagation across countries having different income levels (High-Income, Upper-Middle-Income, Lower-Middle-Income, and Low-Income). The number on the left side and on the right side of income level represents the total number of articles of all those countries that belong to a specific income level. The number in the middle shows the total amount of articles that has propagated the news between two different income levels. We represent each income level with a color to depict the difference between them. The streams of each income level show the propagation across different income levels.

For Global Warming, we observe that news from high-income countries propagated mostly to other high-income countries (blue area on the bottom graph in Fig. 7), with an exception of one news article which propagated to a low-income country (Nigeria). An interesting spreading involves low-income countries, e.g., Iran and Nigeria, where news articles from these countries propagated the news to high-income countries.

For the FIFA World Cup (see Fig. 7), the most frequent countries which appeared to have interesting facts were Germany, Spain, the United Kingdom, and India. News propagated from Germany and Spain to other European countries that have minor economical differences. However, Spain also had some news regarding the FIFA World Cup which propagated news to lower-ranked countries such as India, Bangladesh, Brazil, and South Africa. Most of the articles propagated the news from the United Kingdom to all other economically lower-ranked countries except one article from Germany. The United States appeared to have a mixed combination of articles propagating from and to the United States, Asian and European countries.

For earthquakes, some European countries such as Switzerland, Germany, Austria, France, Portugal and Brazil appeared to have a significant amount of propagation. Apart from many articles that propagated news to other European countries, one article appeared to propagate from Europe to Australia. Overall, there is information crossing economic barriers but we can observe that half of the news articles (yellow and green lines in Fig. 7) related to all events did not cross economical barriers.

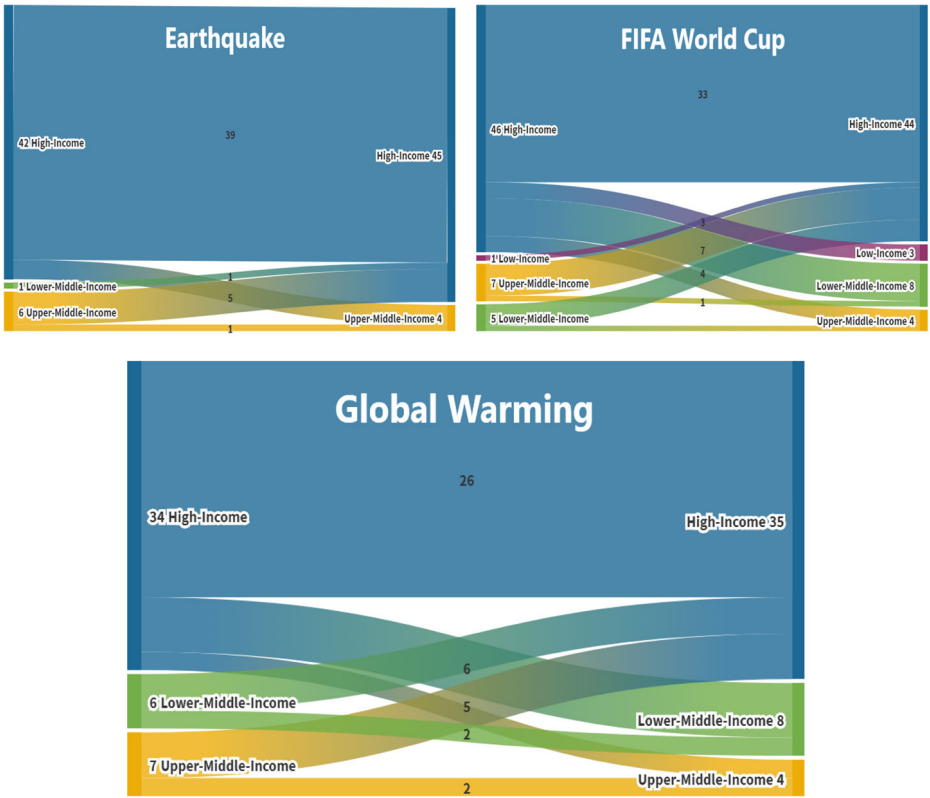

**Fig. 7** Illustration of news spreading across different income levels (from top blue to bottom yellow: High-income, Lower-Middle-Income, Upper-Middle-Income and Low-income) related to earthquakes, FIFA World Cup and Global Warming events, respectively

### 5.2.2 Time zone barriers

To analyze spreading over geo-location, we mapped news articles on Google maps and drew links indicating the total number of propagations among countries to visualize the distance (see Fig. 8). Similar statistics have been shown using a chord diagram - the first three diagrams show propagation among time zones with specific time zone names, the next three with time zone values (see Fig. 9).

We observed several interesting and contrasting effects of different time zone on information propagation on all events. For earthquakes European countries such as Portugal, United Kingdom, Germany, Switzerland, and Brazil surfaced as the most popular in spreading news articles to other countries such as Taiwan, Canada, United States, Australia, and Israel. The difference in time zones among these countries lies between 3 - 13 hours (see Fig. 9). When we looked at FIFA World Cup news articles, the results were somewhat different. Mostly news propagated from the United Kingdom, United States, Spain, and India. The destined countries were mostly European such as Spain, Brazil, United Kingdom, as well as Asian countries such as Bangladesh, India, and Pakistan. The difference in time zone varied. Most of the articles propagated news to countries which had a time difference of 6 or 4 hours.

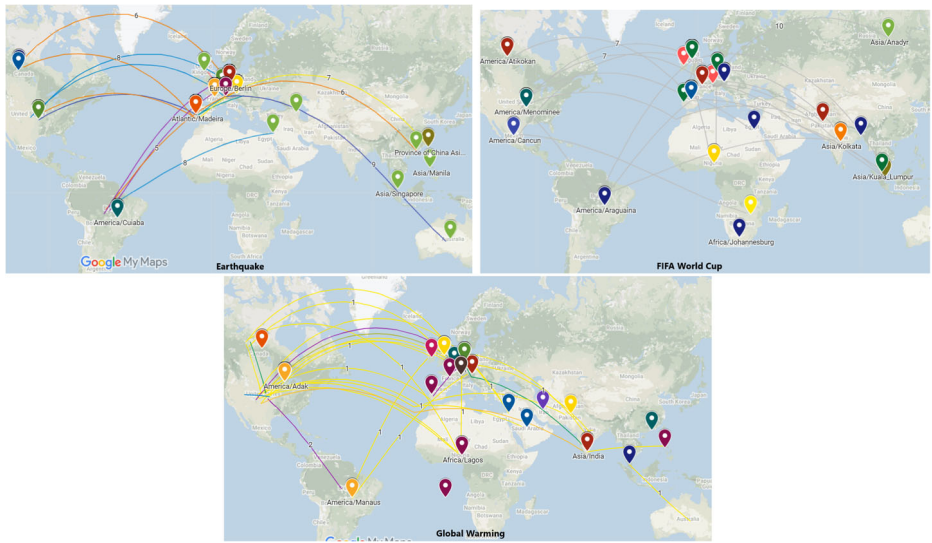

**Fig. 8** Illustration of news propagation on Google Maps across different time zones related to earthquakes, FIFA World Cup and Global Warming

Lastly, we observed interesting factors related to Global Warming. India, Canada, and the United States appeared to have a larger time difference and more articles that propagated news to other countries. Countries that updated/received news from these countries were Saudi Arabia, Nigeria, the Philippines, Belgium, Brazil, Pakistan, Germany, and Switzerland.

### 5.2.3 Geographical barriers

To analyze the geographical impact on news propagation regarding distinctive events, we examined the distribution of publishers and articles over geographical distances among countries. For geographical influence, the distribution of publishers and articles has been illustrated on maps where green, yellow and red color depict the number of instances greater than 10, 50, and 100 respectively (see Fig. 10). In simple words, red, yellow, and green show the significant, medium, and smaller number of articles that are spreading among countries. Initially, our dataset had news articles published over different time periods. When we put them in temporal order based on propagation, the number of articles was reduced. The geographical distribution of publishers related to different events can show the significance of a country. Similarly, we can compare the distribution of news articles with an area or a country. Firstly, we categorized the articles' and publishers' distribution as high (green color), medium (yellow color), low (red color), and none (grey color). The low category indicates less than or equal to 10 publishers and articles whereas medium and high indicated less than or equal to 50 and 100 publishers, respectively. News publishers related to Global Warming are mostly from the United States, United Kingdom, and United Arab Emirates (145, 59, and 175 respectively) as can be seen in green (high category) in Fig. 10. All the countries (Ireland, Nigeria, Spain, Australia) lie in the medium category except ten countries with low category. There is only one country, the United States, which fell into the high category, with

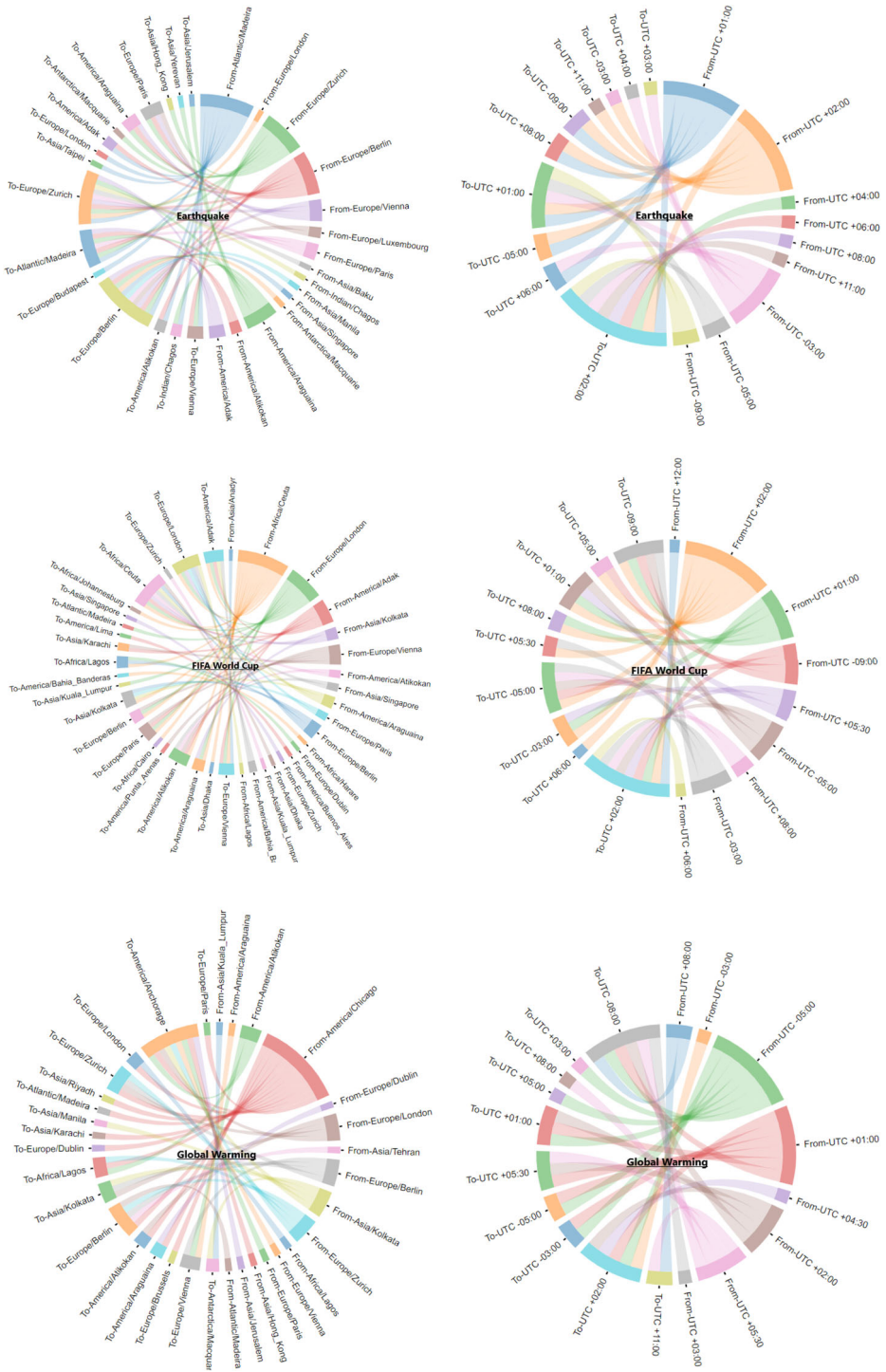

**Fig. 9** Propagation depiction across different time zones illustrated with both a difference among time zone locations and UTC time

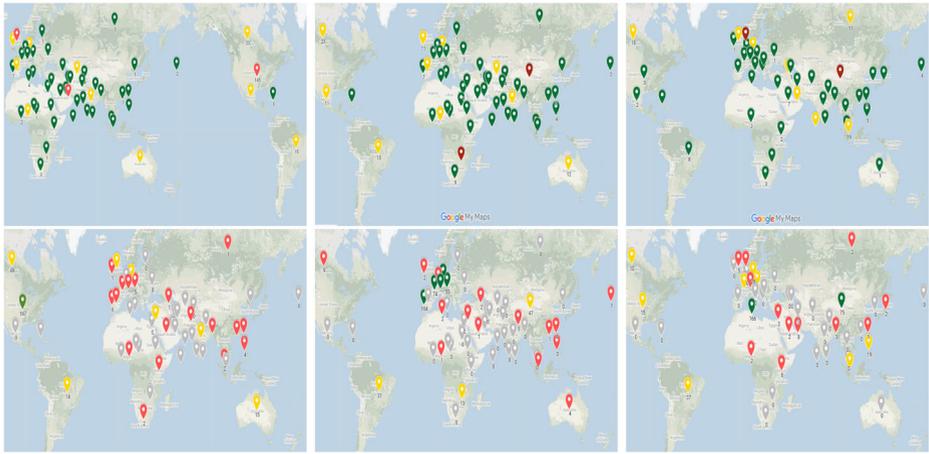

**Fig. 10** An illustration of publishers' (first row) and articles' (second row) distribution over Google Maps for Global Warming, earthquakes, and FIFA World Cup events from left to right, respectively. Red, yellow and green colors show the significant, medium, and smaller amount of articles or news publishers

167 articles when considering the articles' distribution. 35 countries out of 55 happened to have less than two articles, whereas only 8 countries existed with the double-figure values.

Looking at the geographical distribution for the earthquake domain, we see that only two countries - the United Kingdom and the United States, had a high number of publishers (see Fig. 10). For low and medium categories, there were random distributions over the map, however the low category comprises of as many as double than those in the medium category. Figure 10 shows the distribution of news articles related to earthquakes. Most prominent pins are of grey color which means that they did not propagate news. There were five countries nearby pointing to the high category. These countries were Austria, France, Portugal, Switzerland, and Germany with 55, 74, 104, 185, and 181 articles, respectively. Countries in the medium category (the United Kingdom, the United States, and Brazil) also have huge differences geographically. Figure 10 previewed the publisher's distribution regarding the FIFA World Cup. Mostly news publishers belonged to UK and US with a count of 134 and 109. Eight countries (Australia, British Indian Ocean Territory, Canada, Germany, Pakistan, Portugal, Switzerland, and UAE) stood in the medium category with a count of 13, 26, 18, 17, 11, 19, 11, 19, respectively. Finally, related to the FIFA World Cup, most articles were published in Spain (166) and U.S. (75). Austria, Brazil, France, Nigeria, Taiwan, Canada, Germany, Portugal, had news articles in the medium category (with a count of 11, 37, 13, 19, 15, 10, 28, 16, news articles respectively). Overall, our studies depicted that countries which had large geographical area have more publishers for earthquakes than the FIFA World Cup and Global Warming.

### 5.2.4 Political barriers

Primarily, each publisher resides in one of the sixteen classes as displayed in Table 6. According to the table, events related to earthquakes were only published by those publishers which were politically neutral, progressive, and impartial. Publishers having anti-communist, pluralism, and new-left political ideals were more toward spreading news related to Global Warming. The other 10 categories show a presence in all events (see

Fig. 11). Since there is lack of data about political alignments of other publishers, therefore it is difficult to infer more interesting relations between political alignments and events.

### 5.2.5 Cultural barriers

Since the representation of culture has already been described with 6 dimensions (see Table 2), we found that most of the countries which appeared to spread news related to all events through their cultural dimensions were different from each other. For instance, Argentina is the only country that appeared to propagate news related to the FIFA World Cup. Argentina has a moderate score in each culture dimension in comparison to its long term orientation. It has a score as low as 20. We divide the list of cultural values into four categories: low ($< 30$), upper lower ($>= 30\ \&\ < 55$), upper higher ($>= 55\ \&\ < 80$), and high ($>= 80$) and compare the cultural values among countries and observe the spreading patterns in each event. Figure 12 illustrates information propagation between the countries with different power distances (PDI). We can see that countries with low category only propagate news to those countries that stand in the upper-lower and upper-higher categories (indicated with green lines), whereas countries with high category only propagate news to those with upper-higher category (indicated with red lines).

A list of the countries which draw attention to Global Warming include India, United States, Germany, Switzerland, Canada, Portugal, the United Kingdom, and Brazil whereas the list of countries underlined for the earthquake event are United States, Switzerland, Germany, United States, Portugal, Australia, Canada, Brazil, Israel, Taiwan, Armenia, Hong Kong, and the Philippines. Furthermore, for the FIFA World Cup, countries propagate more news articles were Singapore, United States, Russia, Brazil, United States, India, Spain, United States to Canada, Pakistan, Austria, India, Spain, Brazil, Canada, Austria, and France. For all three events, some of the countries share culture and also propagate news. Table 7 shows the list of countries which propagate news and share one of the cultural dimensions (Table 8).

**Table 6** Proclivity of news publishers with specific political class toward different events

| No. | Classes of political alignment | Event type |
| --- | --- | --- |
| 1 | Anti-communist | Global Warming |
| 2 | Catholic | Global Warming, Earthquake |
| 3 | Centrism | Global Warming, FIFA World Cup, Earthquake |
| 4 | Conservative | FIFA World Cup, Global Warming |
| 5 | Independent | FIFA World Cup, Global Warming |
| 6 | Liberalism | FIFA World Cup, Global Warming, Earthquake |
| 7 | New Left | Global Warming |
| 8 | Pluralism | Global Warming |
| 9 | Social Liberalism | FIFA World Cup, Global Warming, Earthquake |
| 10 | Left Wing | FIFA World Cup, Global Warming |
| 11 | Center Right | FIFA World Cup |
| 12 | Moderate | FIFA World Cup |
| 13 | Progressive | FIFA World Cup |
| 14 | Impartiality | Earthquake |
| 15 | Progressive | Earthquake |
| 16 | Neutral | Earthquake |

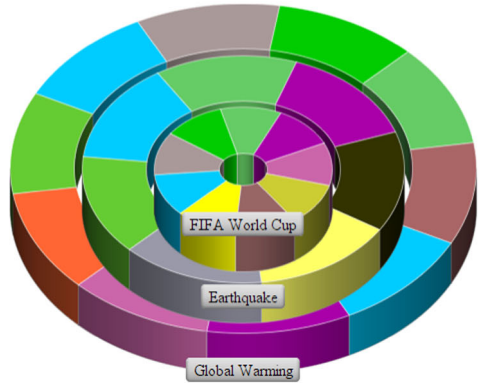

**Fig. 11** Different events with political classes of news publishers

In short, for earthquakes, articles propagate news from Switzerland, Germany, France, Austria. We can see that they have more or less similar culture. There is a fewer number of articles that propagate news toward those countries which are quite different culture-wise such as Switzerland to Australia, Austria to Brazil, and Germany to Canada. For the FIFA World Cup, more articles propagate news to those countries which belong to different culture such as UK to Spain, US to Spain. For Global Warming, similar to the earthquake events, articles propagate news to those countries which are similar in culture.

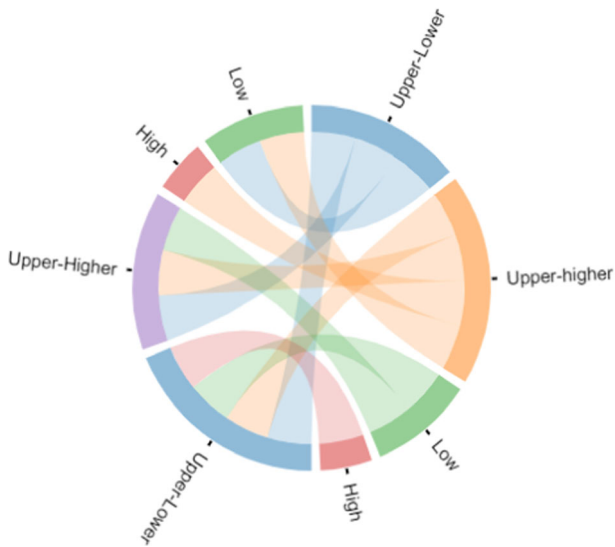

**Fig. 12** News propagation visualization for one cultural dimension (PDI) in Global Warming enlarged from Fig. 13

**Table 7** Table illustrates the list of countries that either share culture or not and propagates news articles

| Event type | News propagation | Dultural dimension | Share culture |
|---|---|---|---|
| Global_Warming | India to the U.S., the U.S. to Brazil. | - | No |
| Global_Warming | the U.S. to France, Portugal to Germany. | - | No |
| Earthquake | Portugal to the U.S., Canada to Germany, Brazil to Israel. |  | No |
| Earthquake | Brazil to Armenia, Portugal to Canada. | - | No |
| Earthquake | Switzerland to Hong Kong, Brazil to Switzerland. | - | No |
| Earthquake | Brazil to Germany, and the Philippines to Germany. | - | No |
| Earthquake | The U.S. to Switzerland, Germany to the U.S. | IDV-MAS, IDV-IVR, MAS-IVR, IDV-MAS. | Yes |
| Earthquake | Switzerland to Australia, Australia to Germany. | IDV-IVR, MAS-IVR, IDV-MAS, MAS-UAI. | Yes |
| Earthquake | Portugal to Taiwan. | MAS-IVR. | Yes |
| FIFA World Cup | Singapore to Canada, Brazil to India, Spain to Canada. | IDV-IVR, PDI-MAS, PDI-LTO,˜IDV-MAS, IDV-LTO, MAS-LTO. | Yes |
| FIFA World Cup | The U.S. to Austria, Singapore to Spain, and Brazil to Spain. | MAS-IVR, LTO-IVR, IDV-IV, PDI-MAS, PDI-LTO, IDV-MAS. | Yes |
| FIFA World Cup | the U.S. to Pakistan, Russia to Austria, the U.S. to Spain. | PDI-MAS, MAS-IVR, MAS-IV, IDV-MAS, PDI-LTO, IDV-MAS. | Yes |
| FIFA World Cup | the U.S. to France, Austria to Canada, and the U.K. to the U.S. | PDI-MAS, MAS-IV, MAS-IV, LTO-IVR, UAI-IVR, UAI-IVR, MAS-LTO. | Yes |

To conclude discussion on the cultural barrier, we can say news related to earthquakes has crossed cultural barriers. For instance, the maximum number of countries with a different score in each dimension propagate news to other countries. Only two cultural dimensions (individualism, indulgent versus restraint) were observed as barriers, as the news seems to propagate mostly to countries which are more or less similar in these two dimensions (see the second and the last circle in Fig. 14). For the FIFA World Cup, we observe that half of the countries with a similar score in dimension of individualism propagate news to each other (see Fig. 15). Similarly, half of the countries appeared to spread news concerning Global Warming topic, with a similar score within the individualism dimension (see Fig. 13).

## 6 Results and discussion

Experiments of the proposed methodology on three types of events have brought some insights regarding information propagation barriers in relation to the type of observed events. The results of mono-lingual information cascading indicate that news related to

**Table 8** Score of each cultural dimension for different countries (For the interpretation of values see Section 2)

| Country | Power distance (PDI) | Individualism (IDV) | Masculinity vs Femininity (MAS) | Uncertainty avoidance by individualism (UAI) | Long and short term orientation (LTO) | Indulgence vs Restraint (IVR) |
|---|---|---|---|---|---|---|
| Africa East | 64 | 27 | 41 | 52 | 32 | 40 |
| Africa West | 77 | 20 | 46 | 54 | 9 | 78 |
| Arab countries | 80 | 38 | 53 | 68 | 23 | 34 |
| Argentina | 49 | 46 | 56 | 86 | 20 | 62 |
| Australia | 38 | 90 | 61 | 51 | 21 | 71 |
| Austria | 11 | 55 | 79 | 70 | 60 | 63 |
| Bangladesh | 80 | 20 | 55 | 60 | 47 | 20 |
| Belgium | 65 | 75 | 54 | 94 | 82 | 57 |
| Brazil | 69 | 38 | 49 | 76 | 44 | 59 |
| Bulgaria | 70 | 30 | 40 | 85 | 69 | 16 |
| Canada | 39 | 80 | 52 | 48 | 36 | 68 |
| Chile | 63 | 23 | 28 | 86 | 31 | 68 |
| China | 80 | 20 | 66 | 30 | 87 | 24 |
| Colombia | 67 | 13 | 64 | 80 | 13 | 83 |
| Croatia | 73 | 33 | 40 | 80 | 58 | 33 |
| Czech Rep | 57 | 58 | 57 | 74 | 70 | 29 |
| Denmark | 18 | 74 | 16 | 23 | 35 | 70 |
| El Salvador | 66 | 19 | 40 | 94 | 20 | 89 |
| Estonia | 40 | 60 | 30 | 60 | 82 | 16 |
| Egypt | 70 | 25 | 45 | 80 | 7 | 4 |
| Finland | 33 | 63 | 26 | 59 | 38 | 57 |
| France | 68 | 71 | 43 | 86 | 63 | 48 |
| Germany | 35 | 67 | 66 | 65 | 83 | 40 |
| Great Britain | 35 | 89 | 66 | 35 | 51 | 69 |
| Greece | 60 | 35 | 57 | 112 | 45 | 50 |
| Hong Kong | 68 | 25 | 57 | 29 | 61 | 17 |
| Hungary | 46 | 80 | 88 | 82 | 58 | 31 |
| India | 77 | 48 | 56 | 40 | 51 | 26 |
| Indonesia | 78 | 14 | 46 | 48 | 62 | 38 |
| Iran | 58 | 41 | 43 | 59 | 14 | 40 |
| Ireland | 28 | 70 | 68 | 35 | 24 | 65 |
| Italy | 50 | 76 | 70 | 75 | 61 | 30 |
| Japan | 54 | 46 | 95 | 92 | 88 | 42 |
| Korea South | 60 | 18 | 39 | 85 | 100 | 29 |
| Latvia | 44 | 70 | 9 | 63 | 69 | 13 |
| Lithuania | 42 | 60 | 19 | 65 | 82 | 16 |
| Luxembourg | 40 | 60 | 50 | 70 | 64 | 56 |
| Malaysia | 104 | 26 | 50 | 36 | 41 | 57 |

**Table 8**  (continued)

| Country | Power distance (PDI) | Individualism (IDV) | Masculinity vs Femininity (MAS) | Uncertainty avoidance by individualism (UAI) | Long and short term orientation (LTO) | Indulgence vs Restraint (IVR) |
| --- | --- | --- | --- | --- | --- | --- |
| Malta | 56 | 59 | 47 | 96 | 47 | 66 |
| Mexico | 81 | 30 | 69 | 82 | 24 | 97 |
| Morocco | 70 | 46 | 53 | 68 | 14 | 25 |
| Netherlands | 38 | 80 | 14 | 53 | 67 | 68 |
| New Zealand | 22 | 79 | 58 | 49 | 33 | 75 |
| Norway | 31 | 69 | 8 | 50 | 35 | 55 |
| Pakistan | 55 | 14 | 50 | 70 | 50 | 0 |
| Peru | 64 | 16 | 42 | 87 | 25 | 46 |
| Philippines | 94 | 32 | 64 | 44 | 27 | 42 |
| Poland | 68 | 60 | 64 | 93 | 38 | 29 |
| Portugal | 63 | 27 | 31 | 104 | 28 | 33 |
| Romania | 90 | 30 | 42 | 90 | 52 | 20 |
| Russia | 93 | 39 | 36 | 95 | 81 | 20 |
| Serbia | 86 | 25 | 43 | 92 | 52 | 28 |
| Singapore | 74 | 20 | 48 | 8 | 72 | 46 |
| Slovak Rep | 104 | 52 | 110 | 51 | 77 | 28 |
| Slovenia | 71 | 27 | 19 | 88 | 49 | 48 |
| Spain | 57 | 51 | 42 | 86 | 48 | 44 |
| Sweden | 31 | 71 | 5 | 29 | 53 | 78 |
| Switzerland | 34 | 68 | 70 | 58 | 74 | 66 |
| Taiwan | 58 | 17 | 45 | 69 | 93 | 49 |
| Thailand | 64 | 20 | 34 | 64 | 32 | 45 |
| Trinidad and Tobago | 47 | 16 | 58 | 55 | 13 | 80 |
| Turkey | 66 | 37 | 45 | 85 | 46 | 49 |
| U.S.. | 40 | 91 | 62 | 46 | 26 | 68 |
| Uruguay | 61 | 36 | 38 | 100 | 26 | 53 |
| Venezuela | 81 | 12 | 73 | 76 | 16 | 100 |
| Vietnam | 70 | 20 | 40 | 30 | 57 | 35 |

sports and natural disasters propagated smoothly based on the fact that the topic of discourse was contiguous, whereas for Global Warming the conversation was more divergent (see Fig. 4). In other words, after looking into the text within chains/communities of articles, we found that news articles related to Global Warming discuss more about other topics rather than only Global Warming whereas chains of the other two topics are more focused on the relevant description (see Section 5.1.2). The concept of understanding information propagation through information cascading has been addressed only in the context of social networks. We are focused to use the same cascading concept and structure over news articles. Although social media is a attractive and effective way of information spreading, a large number of people still rely on newspapers and have a habit to follow print or broadcast mainstream media. While performing cascading experiments, we mainly came across two challenges: 1) Social networks are based on a well-defined and structured architecture whereas in the case of news, such structures are unavailable, and 2) In many cases social networks provide feature of translation that makes a bit easy to find the textual similarity whereas for news articles such features are still not mature enough.

Analysis of multi-lingual temporal spreading suggests that news related to natural disasters exhibits relevant and longer cascading chains of articles indicating smoother information propagation than in the domain of sports and climate changes. We conclude two types of results at the end of this analysis. Firstly, influence of language out of five languages has been identified for each distinctive domain (English language appeared to have more influence for the FIFA World Cup and Global Warming events, whereas Spanish appeared to have more influence for the FIFA World Cup and earthquakes). Secondly, we looked into the text of chains of news articles similar to the mono-lingual analysis and found that news articles related to natural disasters and FIFA World Cup include the relevant communication in long/short chains when compared to the topic of Global Warming (see Fig. 6). Overall, the results of our linguistic analysis correlated with our research hypothesis. Visualizing the temporal information spreading across different languages is one of the challenges in visual analytics. We built a prototype to help understanding temporal spreading though, there are still many improvements required. Firstly, visualization tool is not fine grained at hour, days, week levels. Secondly, 3D tool with zooming functionality can improve this visualization. The analysis of an economic barrier that has been performed over Google Maps using economic rankings of publishers' countries has not established compelling outcomes until we adopted another method. Using the economic ranking of countries, we observed information propagation from economically strong countries to economically weaker countries and vice versa. In all three distinctive domains, news related to Global Warming did not reach economically weaker countries, such as Iran and Nigeria. Using the income-level of countries, we analyzed the spread of news through alluvial diagrams and found that more information was spreading among economically stronger countries than in economically weaker countries (see Fig. 7). Economic barrier is normally more valuable to detect when there is a lot of economic interactivity between two countries and vice versa (Wu, 2007). Other than this, selection of events (conflicting events, popular events) also matters for economic barrier (Segev, 2015). For instance, in our three events, Global Warming is more suitable event for this barrier than FIFA World Cup and Earthquakes.

Our time zone analysis over Google maps suggests a harmony of spreading information related to natural disasters and sports (see Figs. 8, and 9). For natural disasters, European countries appeared to be the origin of the information spreading. In the case of sports, our results were inverse; the origin of the news were mostly countries with larger time zone differences but spreaders were close with respect to time zones (European countries). For climate change, the time zone barrier does not show any significant findings.

Geographical analysis portrays that a high number of publishers and news articles are from the countries that have large geographic areas (see Fig. 10). Apart from this, we did not find any significant differences in process of information spreading among the three distinctive domains. As a result, we could not confirm our research hypothesis on different spreading patterns that were generated for geographical barrier based on three distinctive domains.

Cultural differences also do not demonstrate a correlation with our research hypothesis (see Figs. 13, 14, and 15). Nevertheless, our analysis has shown an interesting observation that Argentina is the country which culturally supports sports activities more than other countries. In fact, we observe that some articles propagate news from a source country to the destination country however it also appeared that countries with common culture propagate news between each other. As a separate result of each domain, we see that news articles related to either Global Warming or earthquakes propagate news to countries that share cultural characteristics whereas news articles related to FIFA World Cup propagate news to those countries with entirely different culture.

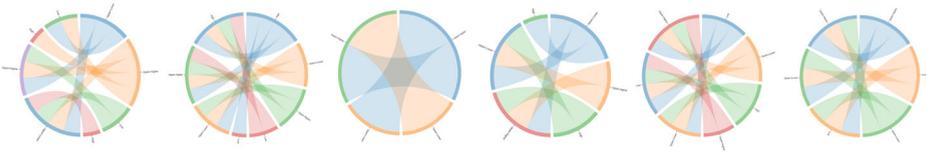

**Fig. 13** News propagation depiction across all cultural dimensions (from left to right: PDI, IDV, MAS, UAI, LTO and IVR) related to Global Warming

Political alignment is important strategy that can be used to control the coverage of news in news agencies. Analysis of political involvement suggests that information on natural disasters spread smoothly, as earthquake-related news was mostly published by publishers that have neutral, progressive, and impartial political alignment (see Table 6). For Global Warming and FIFA World Cup we were not able to summarize the results due to lack of information. Coverage of these both events is not performed by a particular type of publishers (see Table 6).

Overall, our analytical findings suggest that most of the barriers (linguistic, economical, time zone, and political) by and large support our hypothesis. The generalizability of the results is to some extent limited by all the steps including computation and representation. By finding associations we are recognizing that the same information have presence across the barriers and we are not speculating on the cause of that. For instance, if two articles in different languages are similar that we conclude the information cross the linguistic barrier. We notice that the reasons for crossing that can be different and we are not investigating them. Firstly, by changing the cross-lingual similarity method, the results could be clearer and efficient. Secondly, with the availability of publishers' profiles along with the accurate location of their headquarters, the results would be more accurate. Additionally, more experiments on different types of events would enable more precise and robust comparison of information barriers.

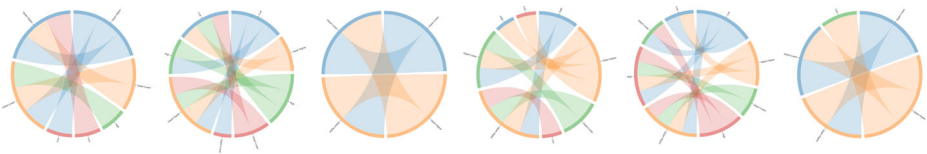

**Fig. 14** News propagation depiction across all cultural dimensions (from left to right: PDI, IDV, MAS, UAI, LTO and IVR) related to earthquakes

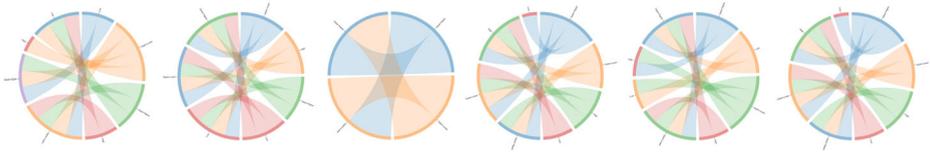

**Fig. 15** News propagation depiction across all cultural dimensions (from left to right: PDI, IDV, MAS, UAI, LTO and IVR) related to FIFA World Cup

## 7 Conclusions

In this paper, we focused on the analysis of information spreading barriers by observing different aspects of news spreading in a global setting. Our motivation was primarily to understand the multilingual information cascading regarding different types of events within news articles, as it is not only valuable for journalists but it is also beneficial in a pragmatic sense for those who wish to follow globally influential-events (e.g. football and basketball in Europe) and investigate the influence of multiple barriers (e.g. economical, geographical, time zone, political and cultural) across different types of events. We firstly characterized the concept of information cascading on news articles in different languages and then find the total and largest cascading chains across distinct kind of events: Global Warming, FIFA World Cup and earthquakes. We also performed analysis to detect the influence of multiple barriers on event-centric news spreading.

In order to answer the first research question of this study - What are the properties (ratio and size) and values of cascading chains in events of different domains? - we identified the number of total communities showing information spreading as 196, 147, and 95 in the earthquake, FIFA World Cup, and Global Warming, respectively. Similarly, the largest cascading chains were in the earthquakes, FIFA World Cup, and Global Warming in English (28 articles), German (24 articles), and English (21 articles), respectively.

However, regarding the second research question - Do the different information cascading chains have any relations with each other? - There is a strong relation between the size and time of cascading chains and information spreading as the time duration of the longest chain of the earthquake was of 2.5 years and 3 months and single day for the FIFA World Cup and Global Warming respectively. Overall, it shows that more news propagated earthquakes than FIFA World and Global Warming.

In order to answer the third research question - Do the economic, geographical, time zone, and cultural values influence event-centric news spreading?- We observed all barriers having a certain effect on news propagation related to events. Firstly, it appears that spreading across languages was influenced by the scope of the event, the geographical size of an area directly related to the amount of news published from this area, places having the same culture publish similar news, news spreads firstly towards areas with adjacent time zones and economic barriers show more news spread upwards like low-income countries to high-income countries. Secondly, a comparison among the events shows that news related to FIFA World Cup propagated toward countries with shorter time zone differences and news propagated between countries with larger time zone differences in other events, economic values indicate that news related to Global Warming did not cross the economic barrier, countries with larger areas have more publishers but European countries have a large number of news articles related to earthquakes, and news related to earthquakes cross cultural barriers effectively than FIFA World Cup and Global Warming.

Finally, the answer to the last research question - What is the correlation of news spreading among events of different domains and political alignment of news publishers? - Was easier to predict based on the simple political alignment of publishers and it suggested that news related to an earthquake event is propagated mostly by those publishers which were politically neutral. Overall, our findings suggest that news published in news articles propagates to a greater degree across languages for natural disasters (earthquake events) than climate change (Global Warming) and sports (FIFA World Cup). In our experiments, we observed more cascading chains as well as longer cascading chains for earthquakes. If we look at the temporal difference between news articles, it shows that there is a larger time difference in the longest cascading chain of earthquake events (which is almost 2.5 years compared to 3 months for FIFA World Cup and 1 day for Global Warming). Our analysis of an economical barrier suggest that for Global Warming events, more news has propagated among economically stronger countries than to economically weaker countries. We found strong evidence that the news related to Global Warming has not crossed the economic barrier to reach Iran and Nigeria, which have currently considered as economically weaker countries. When we looked for information spreading in the domain of natural disaster, cultural barriers were important and difficult to cross. When we moved on to climate change events, linguistic and cultural barrier were difficult to cross, while for sports events, time zone barrier shows to be difficult to cross.

As the result of our research, we have provided a new publicly available dataset to help in understanding information spreading within three domains: natural disasters, climate change and sports. Apart from the data sets, a reusable visualization has been developed to show the real-time spreading of events using cross-lingual news articles. To this end, we utilized Google Maps, alluvial and chord diagrams to look at the spreading of information across the world with economic, cultural, time zone, and geographical distance between larger entities, such as countries. Moreover, we have considered the political involvement of publishers while analyzing the spreading patterns within different domains.

**Acknowledgements** This work was supported by the Slovenian Research Agency under the project J2-1736 Causalify and co-financed by the Republic of Slovenia and the European Union's Horizon 2020 research and innovation programme under the Marie Skłodowska-Curie grant agreement No 812997.

**Data availability** The datasets generated during and/or analysed during the current study are available on the Zenodo repository[15].

# References

Al-Samarraie, H., Eldenfria, A., & Dawoud, H. (2017). The impact of personality traits on users' information-seeking behavior. *Information Processing & Management*, *53*(1), 237–247.

---

[15] https://zenodo.org/record/4117411

Alla, S., Sullivan, S. J., McCrory, P., & Hale, L. (2011). Spreading the word on sports concussion: citation analysis of summary and agreement, position and consensus statements on sports concussion. *British Journal of Sports Medicine*, *45*(2), 132–135.

Andrews, S., Gibson, H., Domdouzis, K., & Akhgar, B. (2016). Creating corroborated crisis reports from social media data through formal concept analysis. *Journal of Intelligent Information Systems*, *47*(2), 287–312.

Bakshy, E., Messing, S., & Adamic, L.A. (2015). Exposure to ideologically diverse news and opinion on facebook. *Science*, *348*(6239), 1130–1132.

Brank, J., Leban, G., & Grobelnik, M. (2017). Annotating documents with relevant wikipedia concepts. *Proceedings of SiKDD*.

Büyüksarıkulak, A. M., & Kahramanoğlu, A. (2019). The prosperity index and its relationship with economic growth: Case of turkey. *Journal of Entrepreneurship, Business and Economics*, *7*(2), 1–30.

Camaj, L. (2010). Media framing through stages of a political discourse: International news agencies' coverage of kosovo's status negotiations. *International Communication Gazette*, *72*(7), 635–653.

Chang, T.-K., & Lee, J.-W. (1992). Factors affecting gatekeepers' selection of foreign news: A national survey of newspaper editors. *Journalism Quarterly*, *69*(3), 554–561.

Cui, Y., Ni, S., Shen, S., & Wang, Z. (2020). Modeling the dynamics of information dissemination under disaster. *Physica A: Statistical Mechanics and its Applications*, *537*, 122822.

Dagon, D., Zou, C. C., & Lee, W. (2006). Modeling botnet propagation using time zones. In *NDSS*, (Vol. 6 pp. 2–13).

Estrada, E. (2011). Community detection based on network communicability. *Chaos: An Interdisciplinary Journal of Nonlinear Science*, *21*(1), 016103.

Glavaš, G., Franco-Salvador, M., Ponzetto, S. P., & Rosso, P. (2018). A resource-light method for cross-lingual semantic textual similarity. *Knowledge-Based Systems*, *143*, 1–9.

He, M., & Lee, J. (2020). Social culture and innovation diffusion: a theoretically founded agent-based model. *Journal of Evolutionary Economics*, 1–41.

Hoftede, G., Hofstede, G. J., & Minkov, M. (2010). *Cultures and organizations: software of the mind: intercultural cooperation and its importance for survival*. McGraw-Hill.

Hong, X., Yu, Z., Tang, M., & Xian, Y. (2017). Cross-lingual event-centered news clustering based on elements semantic correlations of different news. *Multimedia Tools and Applications*, *76*(23), 25129–25143.

Jin, H. (2017). Detection and characterization of influential cross-lingual information diffusion on social networks. In *Proceedings of the 26th International Conference on World Wide Web Companion* (pp. 741–745).

Khosrowjerdi, M., Sundqvist, A., & Byström, K. (2020). Cultural patterns of information source use: A global study of 47 countries. *Journal of the Association for Information Science and Technology*, *71*(6), 711–724.

Krajewski, R., Rybinski, H., & Kozlowski, M. (2016). A novel method for dictionary translation. *Journal of Intelligent Information Systems*, *47*(3), 491–514.

Kumar, S., Saini, M., Goel, M., & Panda, B.S. (2020). Modeling information diffusion in online social networks using a modified forest-fire model. *Journal of Intelligent Information Systems*, 1–23.

Leban, G., Fortuna, B., Brank, J., & Grobelnik, M. (2014). Event registry: learning about world events from news. In *Proceedings of the 23rd International Conference on World Wide Web* (pp. 107–110).

Maurer, P., & Beiler, M. (2018). Networking and political alignment as strategies to control the news: Interaction between journalists and politicians. *Journalism Studies*, *19*(14), 2024–2041.

Miritello, G., Moro, E., & Lara, R. (2011). Dynamical strength of social ties in information spreading. *Physical Review E*, *83*(4), 045102.

Quezada, M., Pe na-Araya, V., & Poblete, B. (2015). Location-aware model for news events in social media. In *Proceedings of the 38th International ACM SIGIR Conference on Research and Development in Information Retrieval* (pp. 935–938).

Raghavan, U. N., Albert, R., & Kumara, S. (2007). Near linear time algorithm to detect community structures in large-scale networks. *Physical Review E*, *76*(3), 036106.

Segev, E. (2015). Visible and invisible countries: News flow theory revised. *Journalism*, *16*(3), 412–428.

Segev, E., & Hills, T. (2014). When news and memory come apart: A cross-national comparison of countries' mentions. *International Communication Gazette*, *76*(1), 67–85.

Şenel, L. K., Yücesoy, V., Koç, A., & Çukur, T. (2017). Measuring cross-lingual semantic similarity across european languages. In *2017 40th International Conference on Telecommunications and Signal Processing (TSP)*, IEEE (pp. 359–363).

Sittar, A., Mladenić, D., & Erjavec, T. (2020). A dataset for information spreading over the news. In *Proceedings of the 23th International Multiconference Information Society SiKDD*, (Vol. C pp. 5–8).

Vulic, I., & Moens, M.-F. (2014). Probabilistic models of cross-lingual semantic similarity in context based on latent cross-lingual concepts induced from comparable data. In *Proceedings of the 2014 Conference on Empirical Methods in Natural Language Processing (EMNLP 2014)* (pp. 349–362). East Stroudsburg: ACL.

Watanabe, K., Ochi, M., Okabe, M., & Onai, R. (2011). Jasmine: a real-time local-event detection system based on geolocation information propagated to microblogs. In *Proceedings of the 20th ACM international conference on Information and knowledge management* (pp. 2541–2544).

Wei, H., Sankaranarayanan, J., & Samet, H. (2020). Enhancing local live tweet stream to detect news. *GeoInformatica*, 1–31.

Wilke, J., Heimprecht, C., & Cohen, A. (2012). The geography of foreign news on television: A comparative study of 17 countries. *International Communication Gazette*, *74*(4), 301–322.

Wu, H. D. (2007). A brave new world for international news? exploring the determinants of the coverage of foreign news on us websites. *International Communication Gazette*, *69*(6), 539–551.

Wu, H. D. (1998). Investigating the determinants of international news flow: A meta-analysis. *Gazette (Leiden, Netherlands)*, *60*(6), 493–512.

**Publisher's note**  Springer Nature remains neutral with regard to jurisdictional claims in published maps and institutional affiliations.